\documentstyle[aps,prl,epsfig,multicol,times]{revtex}
\makeatletter
\newlength\abovecaptionskip \newlength\belowcaptionskip
\def\@makecaption#1#2{%
 \vskip\abovecaptionskip \sbox\@tempboxa{#1: #2}%
 \ifdim \wd\@tempboxa >\hsize #1: #2\par \else \global \@minipagefalse
 \hb@xt@\hsize{\hfil\box\@tempboxa\hfil}%
 \fi \vskip\belowcaptionskip} \makeatother

\begin{document}
\begin{multicols}{2}
{\Large \bf \noindent Comment on "Scaling of Atmosphere and Ocean Temperature Correlations in
Observations and Climate Models"}
\draft

\bigskip
In a recent Letter \cite{Fraed}, Fraedrich and Blender (FB) studied the scaling of
atmosphere and ocean temperature. They analyzed the fluctuation functions $F(s)\sim s^{\alpha}$ of monthly temperature records (mostly from grid data) by using the  detrended fluctuation analysis (DFA2) and claim that the scaling exponent $\alpha$  over the inner continents is equal to $0.5$, 
being characteristic of uncorrelated random sequences.
Here we show that this statement is (i) not supported by their own analysis and (ii) disagrees with the analysis of the daily observational data from which the grid monthly data have been derived.
We conclude that also for the inner
continents, the exponent is between $0.6$ and $0.7$, similar as for the
coastline-stations.

(i) Figure 1a in [1] shows the representative results of FB for $F(s)$ for the inner continental site of Krasnojarsk (observational, grid and model data). Close inspection of the curves shows that none of the observational data approaches the exponent 0.5, but rather yield exponents close to 0.6 or above.

\begin{figure}\centering
\epsfxsize8.5cm\epsfbox{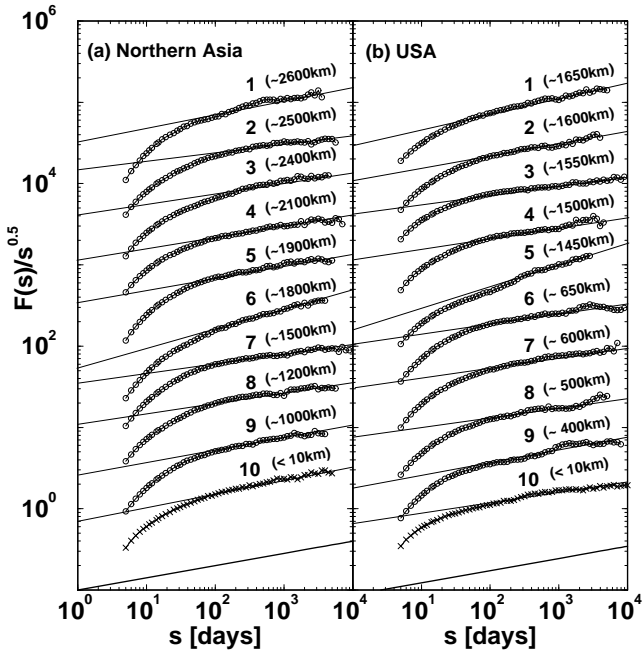}
\parbox{8.5cm}{\caption[]{\small
Results of DFA2 for the daily maximum temperature
records of 18 inner continental stations ($\circ$) and two coastline stations 
($\times$): (a) 1 Urumchi, 2 Tomsk, 3 Atbasar, 4 Chita, 5 Olekminsk, 6 Horog, 
7 Swerdlowsk, 8 Surgut, 9 Jakutsk, 10 Aleksandrovsk; (b) 1 Huron, 
2 Academy, 3 Cheyenne, 4 Gothenburg, 5 Gunnison, 6 Spokane, 7 Winnemucca, 
8 Pendleton, 9 Tusc>on, 10 New York. The estimated minimum distance 
to the oceans is written in parentheses.
 The scale of $F(s)$ is arbitrary. The 
straight lines in the curves represent the  best fits between $s=150$ and 2500. 
 For each curve, the variance of the slope is
about $0.01$.
The line at the 
bottom has a slope of $0.15$, corresponding to $\alpha=0.65$.}
\label{fig:1}}\end{figure}

(ii) Figure 1 shows representative results of $F(s)$ (DFA2) obtained from daily observational records, for 18  inner continental sites 
in North America and Asia, where the stations are between 400 and 2600 km 
away from the ocean.  For comparison, we also show the results for two coastline stations. The maximum
s-value in each curve is below one quarter of the length of the corresponding record.
 To facilitate the evaluation of the data, we have divided $F(s)$ by $s^{0.5}$. A plateau now indicates loss of correlations. 
One can see clearly that none of the curves approaches a plateau, i. e. an 
exponent $\alpha = 0.5$ is never seen. All  asymptotic slopes have values above $0.6$. The mean value of $\alpha$, averaged over all innercontinental stations, is 0.65 $\pm 0.04$. There is no remarkable difference between 
coastline and inner continental stations. Figure 2 shows the dependence of 
$\alpha$ on the shortest distance to the oceans. There is {\it no} tendency 
towards a lower exponent at larger distances.

\begin{figure}\centering
\epsfxsize7.6cm\epsfbox{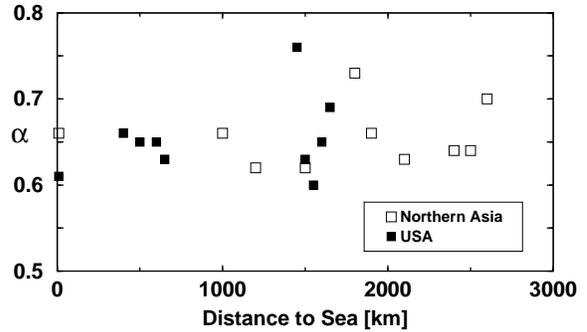}
\parbox{8.5cm}{\caption[]{\small 
The scaling exponent $\alpha$ (obtained from the slopes of the straight lines in Fig.1) 
as a function of the distance to the oceans. }
\label{fig:2}}\end{figure}

Finally, we like to comment on the claim that $\alpha \simeq 1$ for sea surface
temperatures. As has been shown in \cite{Mon}, there is a remarkable crossover
at about 1 year in 
 the sea surface temperatures: 
 At small 
scales, the exponent is significantly larger than 1, $\alpha \simeq 1.3$, while at 
large time scales $\alpha$ is between $0.65$ and $0.95$, with the average at $0.8$.

This work has been supported by the Deutsche Forschungsgemeinschaft and the
Israeli Science Foundation.\\

\noindent A. Bunde$^{1}$, J. F. Eichner$^{1}$, S. Havlin$^{2}$, E. Koscielny-Bunde$^{1,3}$,
H. J. Schellnhuber$^{3,4}$, and D. Vjushin$^{2}$\\
\noindent$^{1}$ Institut f\"ur Theoretische Physik III, 
Universit\"at Giessen, D-35392 Giessen, Germany\\
\noindent$^{2}$ Minerva Center and Department of Physics,
Bar-Ilan University, Ramat-Gan 52900, Israel\\
\noindent$^{3}$ Potsdam Institute for Climate Impact Research, D-14412 Potsdam, Germany\\
\noindent$^{4}$ Tyndall Center for Climate Change Research, Norwich, NR4 7JD, England


\end{multicols}

\end{document}